# Non-invasive Deep-Brain Imaging with 3D Integrated Photoacoustic Tomography and Ultrasound Localization Microscopy (3D-PAULM)

Yuqi Tang, Zhijie Dong, Nanchao Wang, Angela del Aguila, Natalie Johnston, Tri Vu, Chenshuo Ma, Yirui Xu, Wei Yang, Pengfei Song, and Junjie Yao

*Abstract*—Photoacoustic computed tomography (PACT) is a proven technology for imaging hemodynamics in deep brain of small animal models. PACT is inherently compatible with ultrasound (US) imaging, providing complementary contrast mechanisms. While PACT can quantify the brain's oxygen saturation of hemoglobin ($sO_2$), US imaging can probe the blood flow based on the Doppler effect. Further, by tracking gas-filled microbubbles, ultrasound localization microscopy (ULM) can map the blood flow velocity with sub-diffraction spatial resolution. In this work, we present a 3D deep-brain imaging system that seamlessly integrates PACT and ULM into a single device, 3D-PAULM. Using a low ultrasound frequency of 4 MHz, 3D-PAULM is capable of imaging the whole-brain hemodynamic functions with intact scalp and skull in a totally non-invasive manner. Using 3D-PAULM, we studied the mouse brain functions with ischemic stroke. Multi-spectral PACT, US B-mode imaging, microbubble-enhanced power Doppler (PD), and ULM were performed on the same mouse brain with intrinsic image co-registration. From the multi-modality measurements, we future quantified blood perfusion, $sO_2$, vessel density, and flow velocity of the mouse brain, showing stroke-induced ischemia, hypoxia, and reduced blood flow. We expect that 3D-PAULM can find broad applications in studying deep brain functions on small animal models.

## I. INTRODUCTION

PHOTOACOUSTIC computed tomography (PACT) is a hybrid imaging modality that combines the rich contrast of optical absorption and deep penetration of ultrasound detection [1]–[3]. Hemoglobin, as the most commonly used endogenous absorber, offers excellent PA contrast for visualizing blood vessels and their functions in the brain (**Fig. 1a**). A key application of PACT is mapping oxygen saturation of hemoglobin ($sO_2$), by differentiating the oxygenated and deoxygenated hemoglobin using multi-spectral analysis of PA signals. However, without using exogenous contrast, PACT has difficulty in detecting blood flow, due to the lack of speckle contrast. Moreover, PACT with a partial detection aperture often suffers from the so-called 'limited view problem', in which coherent PA signal generation results in missing structures [4], [5]. These shortcomings of PACT can be mitigated by incorporating ultrasound (US) imaging (**Fig. 1a**). US imaging provides information about tissue anatomy and is sensitive to the blood flow based on either the Doppler effect or the speckle fluctuations induced by the flowing red blood cells. Detecting acoustic signals at similar frequencies, PACT and US imaging are highly compatible, and can be integrated into a single system to offer more comprehensive hemodynamic information than a single modality alone [6]–[9].

Nevertheless, traditional PA and US imaging often use relatively low ultrasound frequency (<10 MHz) to accommodate the large penetration depth (> 1 cm), which results in limited spatial resolution for visualizing deep-tissue hemodynamics. Such limitation is particularly true for non-invasive deep-brain imaging on small animal models, in which the intact skull has strong acoustic attenuation at high frequencies and induces severe acoustic aberration. Recently, super-resolution ultrasound localization microscopy (ULM) has overcome the resolution limit [10]–[12]. Using gas-filled microbubbles as strong echoic tracers, ULM can track the positions of the microbubbles over time, and the super-resolved microvessel images can be reconstructed in the deep brain [13], [14]. Thus, further incorporating ULM into the PA and US imaging system can significantly enhance microvasculature imaging capabilities. Robin *et al.* combined PA and US imaging with a spherical transducer array, and demonstrated contrast-enhanced power Doppler (PD) imaging at a volumetric

This work was sponsored by National Institutes of Health (R21EB027981, R21 EB027304, RF1 NS115581, R01 NS111039, R01 EB028143, R21 EB030072), National Science Foundation CAREER award 2144788; and Chan Zuckerberg Initiative Grant (2020-226178).

Y. Tang and Z. Dong contribute equally to the work.

Y. Tang, N. Wang, N. Johnston, T. Vu, C. Ma, Y. Xu, J. Yao are affiliated with the Department of Biomedical Engineering, Duke University, Durham, NC 27703 USA (email: yuqi.tang@duke.edu, junjie.yao@duke.edu)

Z. Dong and P. Song are affiliated with Beckman Institute for Advanced Science and Technology and the Department of Electrical and Computer Engineering, University of Illinois Urbana-Champaign, Urbana, IL 61801 USA (e-mail: zhijied3@illinois.edu, songp@illinois.edu)

A. del Aguila and W. Yang are affiliated with multidisciplinary Brain Protection Program, Department of Anesthesiology, School of Medicine, Duke University, Durham 27710, NC, USA rate of 100 Hz, which, however, is too slow for ULM [15]. Zhao *et al.* implemented 3D ULM with a multimodal PA and US imaging system based on raster scanning of a 1D linear transducer array (15 MHz central frequency), which would have difficulty imaging through the mouse brain with the skull intact [16]. To date, there has been no demonstration of integrated PACT and ULM with a high volume rate, high spatial resolution, and large imaging depth.

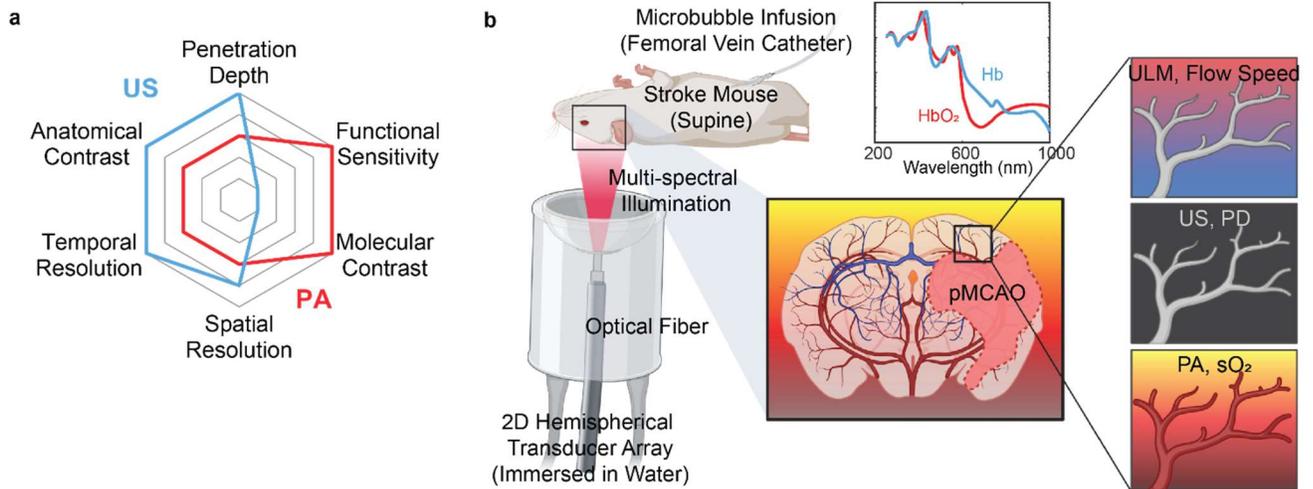

**Fig. 1. 3D integrated photoacoustic tomography and ultrasound localization microscopy. (a)** Radar chart illustrating the imaging capabilities of photoacoustic (PA) and ultrasound (US) imaging. **(b)** Schematic of the imaging system setup. A 2D hemispherical transducer array is immersed in water, with a central aperture for optical fiber insertion. The mouse is placed in a supine position, and the region of interest is positioned around the focal zone of the transducer array (approximately 40 mm from the transducer surface). The system is capable of multi-spectral PA imaging, US B-mode imaging, US power Doppler imaging (PD), and ultrasound localization microscopy (ULM).

In this study, we have developed an integrated imaging system that seamlessly combines 3D PACT and ULM (3D-PAULM). 3D-PAULM works at a relatively low frequency of 4 MHz, which is beneficial for deep brain imaging with both the scalp and skull intact. Besides multi-spectral PACT and ULM, 3D-PAULM can also perform US B-mode imaging and PD imaging, with all the modalities automatically co-registered. We characterized the key performance of 3D-PAULM by numerical simulations and phantom experiments. Then, using an ischemic stroke mouse model, we demonstrated 3D-PAULM for comprehensive functional imaging of the brain, including blood perfusion, microvasculature density, $sO_2$, and blood flow velocity. Our results have highlighted the potential of 3D-PAULM as a non-invasive imaging tool for studying deep brain functions.

## II. METHODS

### A. The Overall 3D-PAULM System

As shown in **Fig. 1b**, the 3D-PAULM system comprises a programable ultrasound scanner (Vantage 256, Verasonics, Kirkland, WA), multiple pulsed lasers for PA excitation, and a customized 2D spherical ultrasound transducer array (Imasonics, France) for both PA and US imaging. The spherical transducer array has 256 elements, with a central opening for inserting an optical fiber bundle. The transducer array has a center frequency of 4 MHz, and −6 dB bandwidth of 45% in the transmit/receive mode, although it was fine tuned to maximize the receiving bandwidth. The transducer array has a curvature radius of 40 mm and total aperture size of 57 mm. The well-sampled field of view (FOV) of the transducer (or the focal zone) is 8×8×8 mm³. During the imaging, the transducer array was facing up inside a temperature-regulated water bath for acoustic coupling. The imaged animal was mounted on an optically and acoustically transparent sample holder at the transducer array's focal zone. The sample holder can be raster scanned by a 2D motorized translation stage for an expanded FOV. The 3D-PAULM system can perform multi-spectral PACT, US B-mode imaging, PD imaging and ULM. The system setup, signal acquisition, and image reconstruction for each imaging modality are detailed below.

### B. 3D US Pressure Field Simulation

Due to the large pitch size of the transducer array (330 µm), synthetic-aperture US imaging was employed to minimize grating lobes [17]. There is a trade-off between the number of individual transmitting elements (Tx elements) and the maximum volumetric frame rate. To investigate the transmitted beam pattern, we conducted a 3D k-Wave simulation and compared the transmission patterns using the 31 Tx elements or 256 Tx elements [18]. For the case of 31-Tx-element, the distance between neighboring Tx elements was 7.5 mm, as shown in **Fig. 2a**. The pixel size in the simulation was 150 µm in x-, y-, and z-directions, with a simulation volume of 62×62×45 mm³. The element geometry in the simulation was consistent with that of the experimental transducer array, i.e., trapezoidal shape with a 40-mm curvature. Forward US transmission simulations were performed, and the maximum pressure field and root-mean-square (RMS) pressure were recorded for each transmission pattern. We recorded the transmitted pressure from a single element (**Fig. 2b**) and compared the accumulated pressure field from the 31-element and 256-element synthetic aperture transmission (**Fig. 2c**).

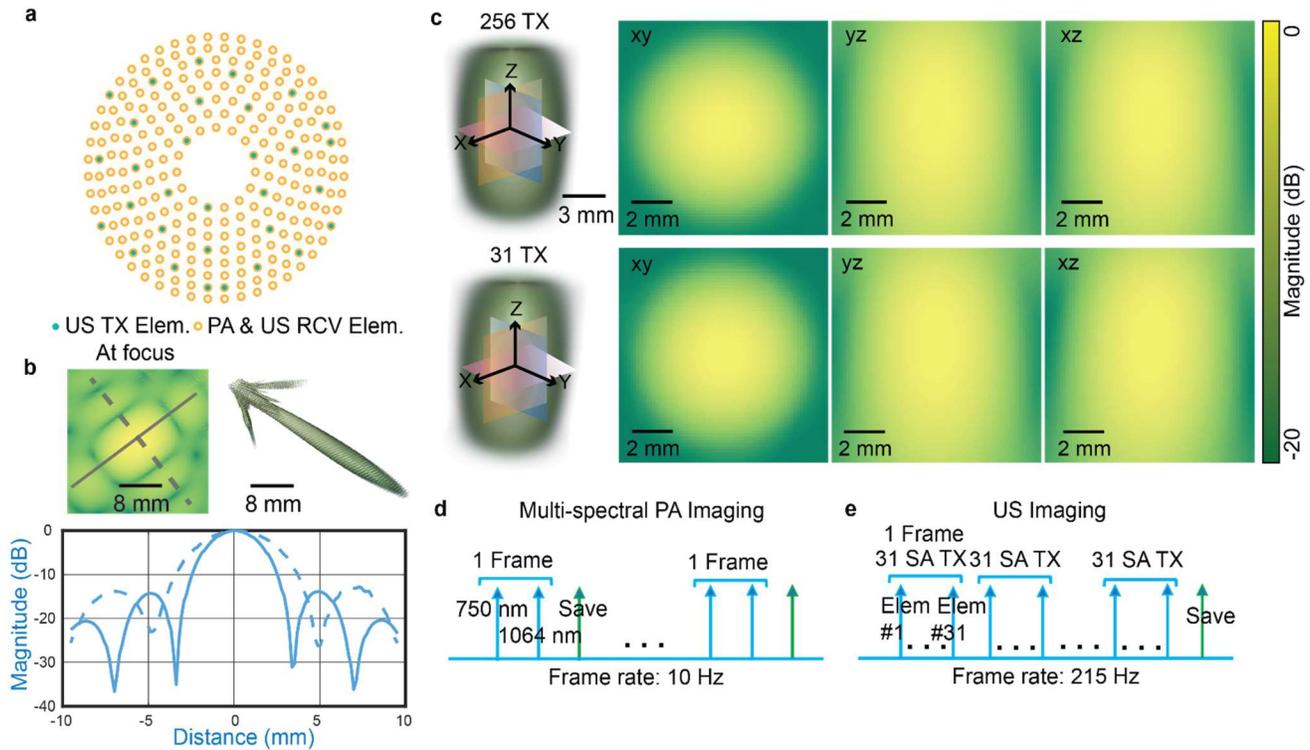

**Fig. 2. System characterization and imaging sequence. (a)** Locations of the transducer elements projected in the lateral plane. Hollow orange circles represent the receiving aperture (256 elements), while solid green circles represent the 31 transmitting elements. **(b)** Simulated pressure field transmitted from a single element at the transducer focus. The pressure profiles along different directions are extracted. **(c)** Simulated pressure field with two different transmission patterns: a 256-element synthetic transmission aperture (left column) and a 31-element synthetic transmission aperture (right column). Pressure is displayed in dB. **(d-e)** Data acquisition sequence of PA and US imaging, with a respective volumetric frame rate of 10 Hz and 215 Hz.

### C. System Characterization

The receiving electrical impulse response of the system was first measured by imaging a thin layer (~80 $\mu$m thick) of gold nanoparticles at 1064 nm. The gold nanoparticles had a diameter of ~80 nm and were highly concentrated, generating a Dirac PA signal. The signal detected by a single element at the center of the transducer array was sampled at 62.5 MHz and averaged over 100 times for noise reduction. Additionally, to characterize the system resolution, a crossed hair target was imaged using both PACT and US B-mode imaging. The hair phantom was placed at the transducer's focal zone, approximately 40 mm from the transducer's surface.

### D. Stroke Animal Preparation

A male C57Bl/6 mouse was used to demonstrate the 3D-PAULM's *in vivo* imaging performance. The animal was 4 months old and had undergone permanent middle cerebral artery occlusion (pMCAO) surgery 72 hours prior to imaging, following the same surgical protocol as described in our previous work [19]. The head was shaved, and the body temperature was maintained at 37°C throughout the experiment. A femoral vein catheter was inserted right before the imaging. The imaging session lasted approximately 10 minutes, including both PA and US data acquisition. Microbubbles were injected via the catheter only for US imaging. Microbubbles were continuously infused using a syringe pump with a concentration of $7\times10^8$ microbubbles/mL and the injection rate was 10 $\mu$L/min. The total volume of microbubbles injected was less than 100 $\mu$L. All the animal experimental procedures were approved by Duke University Institutional Animal Care & Use Committee.

### E. Data Acquisition

The PA and US data were acquired sequentially for the mouse brain. Due to the relatively small focal zone of the 2D transducer array, raster scanning of the mouse by the 2D motorized translation stage was performed to cover the whole mouse brain. A total of 6 scanning positions were used (step size: 3 mm), which results in an extended FOV of 14×11×8 mm$^3$. In PA imaging, three excitation wavelengths at 700nm, 750 nm and 1064 nm were synchronized, and the laser pulse emissions were separated by 200 $\mu$s to form a single multi-wavelength PA frame with a volumetric frame rate of 10 Hz (**Fig. 2d**). A total of 10 PA frames were acquired at each scanning location, with a total acquisition time of 1 second per location.

The US data was acquired after the PA imaging was completed. The synthetic aperture imaging with 31-Tx-element transmission was applied, and the back-scattered echo signals from each transmission event were received with the full aperture, *i.e.,* 256 elements (**Fig. 2e**). A total of 4800 3D US frames were acquired at each scanning position, with a total acquisition time of <3 minutes. To minimize reverberation artifacts, each US transmission event was separated by 150 $\mu$s. The received RF data was batch-transferred to ensure a US volumetric frame rate of 215 Hz.

### F. 3D Multi-spectral PACT

PACT often uses near-infrared light for deep penetration. We first used 1064 nm light (DRL 100, Qutantel Laser) due to its ability to penetrate the entire brain. Two more wavelengths of 700 nm and 750 nm (Surelite OPO Plus, Amplitude Laser) were selected for PA imaging of $sO_2$. All three wavelengths have an 8 ns pulse-width and a pulse repetition rate of 10 Hz. The light was delivered using an optical fiber bundle (Dolan Jenner, 50% coupling efficiency) inserted into the central opening of the 2D spherical array transducer, illuminating the focal zone of the transducer array. The optical fluence on the sample surface at 700 nm, 750 nm, and 1064 nm was 13 mJ/cm$^2$, 15 mJ/cm$^2$, and 75 mJ/cm$^2$, respectively, all of which were within the ANSI limit [20].

At each scanning position, RF data were acquired for 10 laser pulses at each wavelength with a sampling frequency of 20.83 MHz. The RF data from the 10 laser pulses were averaged to improve signal-to-noise ratio (SNR). A 3D delay-and-sum (DAS) method was used for PA image reconstruction, with a voxel size of 60 μm in all dimensions. The reconstructed images for all scanning positions were stitched to form the final 3D whole-brain image.

PA images at 700 nm and 750 nm were used to estimate $sO_2$. To account for frequency-dependent acoustic attenuation in deeper regions and improve the quantification accuracy, the RF data were deconvolved with the electrical impulse response of the transducer array and low-pass filtered at a cutoff frequency of 1 MHz. The relative concentrations of oxygenated and deoxygenated hemoglobin were calculated using the linear unmixing method [21], [22]. We did not perfom optical fluence compensation for the $sO_2$ calcualtion. The optical properties of the brain tissue at 700 nm and 750 nm are similar, allowing for a comparable fluence distribution inside the tissue and thus more accurate calculation of $sO_2$ without fluence compensation.

### G. 3D PD Imaging and ULM

The 3D US B-mode imaging was performed by using the synthetic aperture method, as shown in **Fig. 2a**. The transmitted waveform from each element had a central frequency of 4 MHz, 1 complete burst cycle, and a 10 V peak-to-peak driving voltage. In-phase and quadratic-phase (IQ) volumes were reconstructed using 3D DAS with the same voxel size as PACT (*i.e.*, 60 μm in all dimensions). To mitigate the motion artifacts due to animal breathing, 2D cross-correlation was computed on the lateral maximum intensity projection (MIP) images, using the averaged MIP frame as the reference frame. A correlation threshold of 99% was applied to remove the moving frames, which corresponded to ~10% of total frames. From the US B-mode images, the PD and ULM processing workflow consists of four major steps to reconstruct the brain vasculature images, as shown in **Fig. 3** [23].

Step 1. Spatial-temporal singular value decomposition (SVD) was applied within each block of 600 B-mode volumes to remove clutter signals and extract microbubble signals. The first 150 largest singular values were removed for each block, as demonstrated in **Figs. 3a-b** [24].

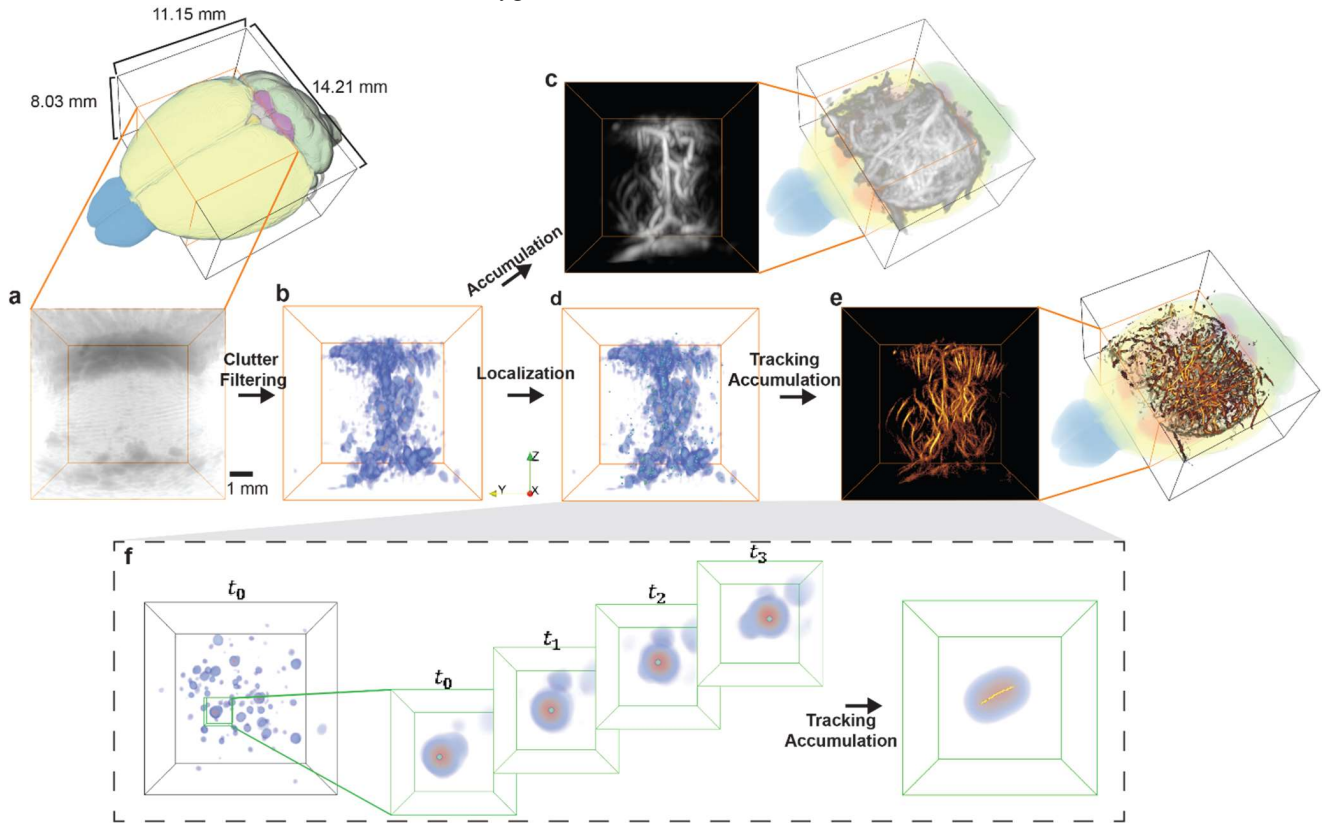

**Fig. 3. ULM processing workflows. (a)** US B-mode images are reconstructed by 3D DAS. **(b)** Spatial-temporal singular value decomposition (SVD) filtering is applied to remove the clutter signals. **(c)** Power Doppler image is formed by accumulating the filtered data. **(d)** Radial symmetry-based localization is performed to extract the positions of microbubble signals (green dots). **(e)** Trajectories retrieved using the 3D tracking algorithm are accumulated to form the ULM density map and flow speed map. **(f)** Illustration of MB localization and tracking.

Step 2. Power Doppler (PD) volumetric images were formed by integrating the SVD-filtered IQ data (**Fig. 3c**). To facilitate microbubble separation, directional filtering and bandpass filtering (20-107 Hz) were used [25].

Step 3. Following the filtering, a radial symmetry-based microbubble localization algorithm was applied to extract the sub-diffraction locations of the microbubbles (**Fig. 3d**) [26].

Step 4. To track the motion of microbubbles across consecutive volumes, a 3D particle tracking algorithm was applied [27]. The microbubble trajectories were then accumulated to generate the ULM density map, as shown in **Fig. 3e**. Additionally, a microbubble flow velocity map was calculated based on the trajectories, providing information on blood flows.

An example of localizing and tracking microbubbles in a water tank is shown in **Fig. 3f**. Similar to PACT, the reconstructed PD and ULM images at all scanning positions were stitched to the final 3D images.

### H. Real-time Image Reconstruction and Display

We utilized a real-time image reconstruction method for multi-spectral PACT that allowed for the immediate display of PA images at all three wavelengths. In DAS, the delay calculation was performed between each element of the array and each pixel in the region of interest (ROI). These delays were then discretized to match the RF signal length, and the corresponding pixel values were added across the transducer array to reconstruct an average initial pressure distribution. To further speed up the DAS process, we employed a GPU-enhanced sparse-matrix-multiplication based method [28].

### I. Vessel Segmentation

Unlike previous ULM techniques that often utilize tens of thousands of frames for microbubble localization, our ULM approach uses only 4800 frames at one scanning position, which may not provide continuous vessels in the final trajectory images [10]. Since vessel continuity is crucial for accurate quantitative analysis of the hemodynamics, we chose to use the 3D PD images for vessel segmentation, which have lower resolution but higher vessel continuity.

The vessel segmentation was performed using the TubeTK library in Python. The 3D PD image of the mouse brain vasculature was first divided into the left and right hemispheres to compare the stroke and healthy regions. To further improve the vessel continuity and connect potentially disconnected vessels, the PD image was repeatedly blurred by 3D Gaussian filters and sharpened by erosion. Then an empirical thresholding was applied to reduce background noise. The preprocessed PD images were subjected to a difference of gaussian blurring to highlight the vessel ridges. A Danielsson distance filter was applied to create a mask for the radii of the vessels. The filtered images were fed to the TubeTK ridge extraction algorithm. About 500 seeds were placed to extract initial vessel-like structures for subsequent discriminant analysis, which defined the image space according to vessel-like properties of curvature, roundness, and ridgeness [29]. The vessel ridges were identified if they have high values of curvature, roundness, and ridgeness. The extracted ridges were saved as a directed graph, which contained the coordinates and radii of the segmented vessels. The vessel length was calculated by integrating the Euclidean distance between all points of each vessel, and the vessel volume was calculated using the length and the radius of the vessels. The final extracted vessel parameters included vessel radius, vessel length, vessel volume, and vessel spatial coordinates. Additional parameters included total vessel length and total vessel number. The segmented individual vessels were also used as spatial masks to extract the functional information from the PA and ULM images.

## III. RESULTS

### A. System Characterization

From the numerical simulations, we found that the spatial pressure fields in the US B-mode imaging were almost identical between the two transmission patterns (31 Tx elements vs. 256 Tx elements), so we used the 31-Tx-element transmission pattern for future experiments to achieve a higher volumetric rate for ULM. The receiving electrical impulse response of the transducer was shown in **Fig. 4a**. The transducer has a central frequency of 4 MHz and −6 dB receiving-only bandwidth of 75%, with slightly better sensitivity towards lower frequencies. It was assumed that all 256 elements had identical impulse responses, and the measured impulse response was used for RF data deconvolution prior to PA image reconstruction. Using the extracted 3D intensity profiles for both PA and US images of the crossed hair phantom, the measured lateral and axial resolution were 260 $\mu$m and 390 $\mu$m for PA, and 270 $\mu$m and 440 $\mu$m for US, respectively (**Fig. 4b**). US imaging has slightly worse lateral resolution than PA imaging is majorly because the transducer was optimized to maximize the receiving sensitivity and bandwidth. With relatively homogeneous light illumination, the effective field of view of the system was approximately 8×8×8 mm³ for both PA and US imaging (**Fig. 3c**).

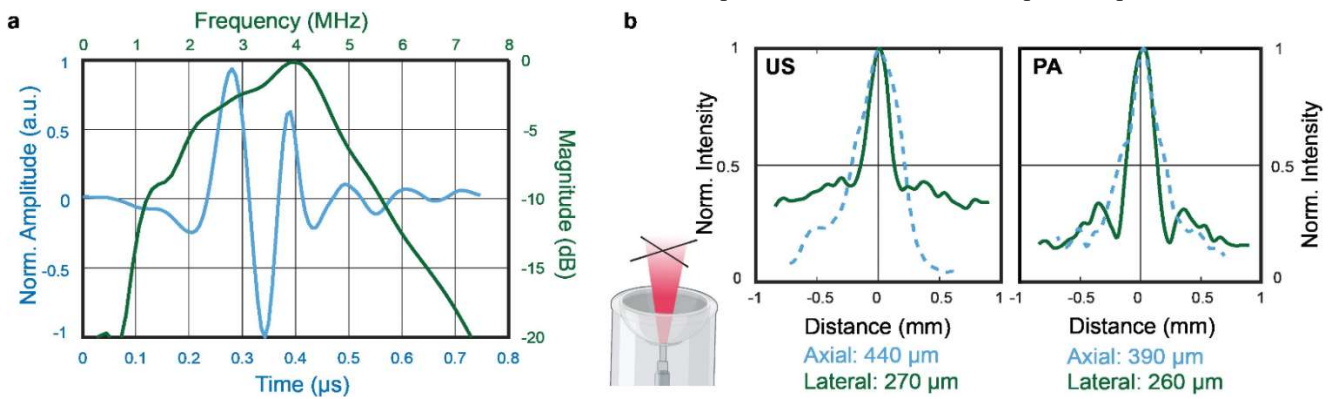

**Fig. 4. System characterization. (a)** The system's receiving electrical impulse response and bandwidth, extracted from PA signals of a nanoparticle sheet. **(b)** PA and US imaging resolutions measured with a crossed hair phantom placed at the transducer focus.

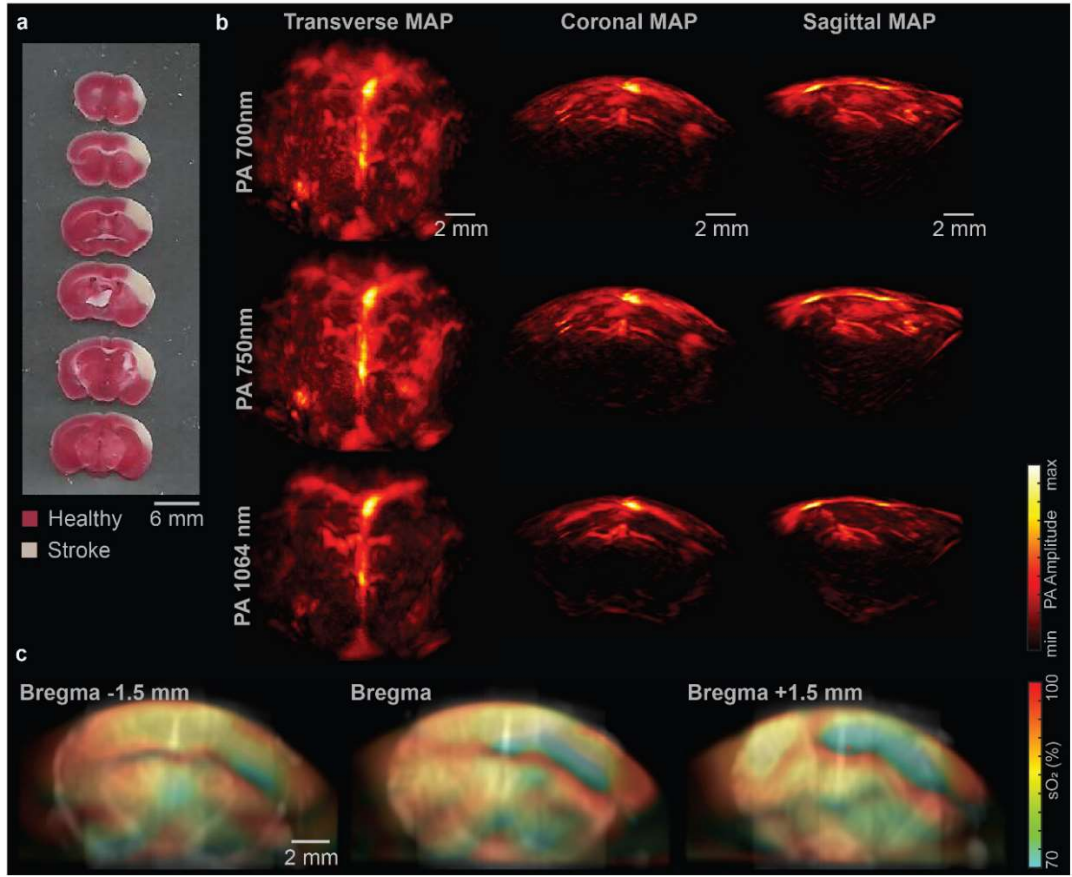

**Fig. 5. Multi-spectral PACT of mouse brain with ischemic stroke.** **(a)** TTC stained 1-mm-thick coronal slices of mouse brain (white area indicates stroke-induced infarct). **(b)** Maximum amplitude projection (MAP) from the transverse, coronal, and sagittal planes of the 3D PA images at 700 nm, 750 nm, 1064 nm. **(c)** $sO_2$ images of the coronal brain slices with a thickness of 1.5 mm.

### B. 3D PA Imaging of Stroke Mouse

After the imaging session of the stroke mouse, we performed histology validation using 2,3,5-triphenyltetrazolium chloride (TTC) staining, which confirmed that the pMCAO procedure primarily affected the cortex region of the right hemisphere in the mouse (**Fig. 5a**). The impact on the hippocampus region was not as evident as in the cortex region.

The PA images of the mouse brain at 1064 nm, 750 nm, and 700 nm are shown in **Fig. 5b**. In the PA images at all three wavelengths, there is a clear reduction in the PA signal intensity in the right hemisphere compared to the left hemisphere. This PA signal reduction is primarily due to the decrease in blood perfusion, and some remaining signals in the stroke region may come from the static blood clots, which may not be detected by the PD imaging or ULM that rely on the flowing blood (shown below).

To demonstrate the functional imaging capability of the imaging system, $sO_2$ maps were extracted from coronal slices with a thickness of 1.5 mm at various positions (**Fig. 5c**). We would like to point out that high-frequency PA signals were more attenuated by tissues and more sensitive to the tissue's optical and acoustic heterogeneity, and thus in this work low-frequency signals (<1 MHz) were used in the $sO_2$ quantification. Moreover, we have observed that the region surrounding the stroke area in the right hemisphere had relatively high $sO_2$, which was not observed in the left hemisphere, as indicated by the white arrows in **Fig. 5c**. It is possibly due to the overcompensation by the collateral blood flow from the left hemisphere [30].

The $sO_2$ maps were consistent with the stroke-induced tissue infarct in the TTC staining results, as reflected by the lower $sO_2$ levels in the cortex and hippocampus regions of the right hemisphere. The pMCAO surgery significantly reduced blood flow in the cortex regions and some of the deeper regions. In addition, the cortical region of the left hemisphere around Bregma −1.5 mm was also slightly affected, which was not clear on the TTC results but clearer on the PD and ULM results shown later. We also observed that the $sO_2$ in the cortex region at Bregma −1.5 mm was higher than those at Bregma +1.5 mm. This suggests that the initially impacted region by the pMCAO surgery may be closer to the Bregma +1.5 mm.

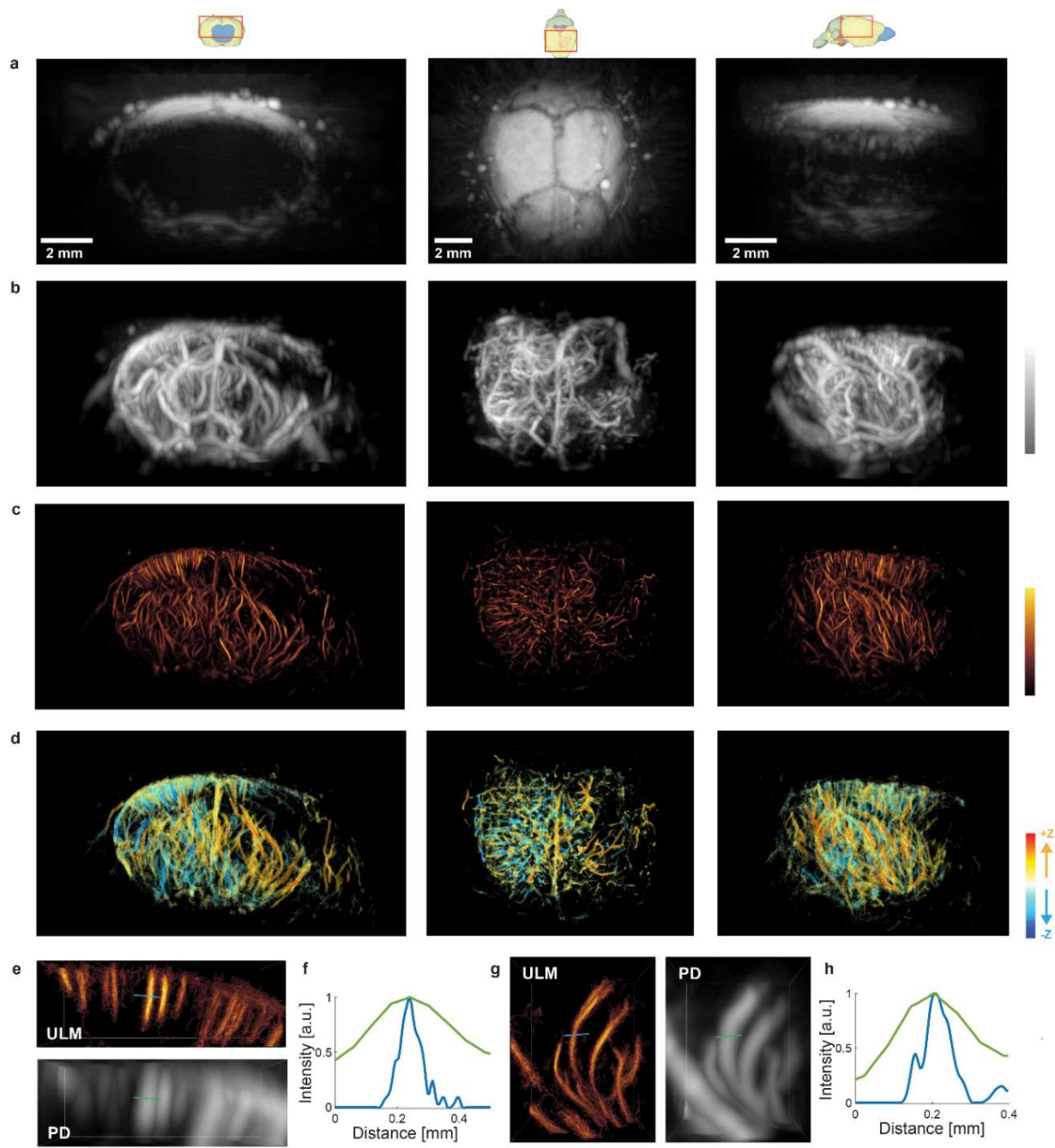

**Fig. 6. 3D US imaging of the mouse brain.** Coronal, transverse, and sagittal view of the US B-mode image **(a)**, US PD image **(b)**, ULM image **(c)**, and the velocity map **(d)**. **(e)** and **(g)** show close-up 3D images of two distinct regions of the PD and ULM density map. **(f)** and **(h)** depict representative vessel profiles by ULM and PD, respectively, illustrating the different spatial resolution.

## C. 3D PD imaging and ULM of Stroke Mouse

While the US B-mode images mainly show the mouse skull due to its strong acoustic reflection (**Fig. 6a**), the PD and ULM images clearly show a reduction in vessel density at the right hemisphere from both the transverse and coronal projections (**Figs. 6b-c**). Both the PD and ULM images show a significant decrease in blood perfusion in the stroke region compared to the healthy region, from both the vessel density and velocity maps. It is worth noting that the PD and ULM images of blood vessels are based on the flow contrast, which requires the actual motion of the blood and the microbubbles. This contrast mechanism is different from the PA imaging of blood vessels that does not require the flow. Thus, the PD and ULM images may not detect the static blood clots formed in the stroke region. On the other hand, PD and ULM images clearly show the diving vessels in the mouse cortex, including the penetrating pia arteries and veins. These diving vessels are not visible in the PA images, mainly due to the aforementioned limited-view problem. Moreover, by tracking the relative motion of microbubbles, ULM can also report the blood flow velocities in absolute units (**Fig. 6d**), which is not available in either the PA or PD imaging.

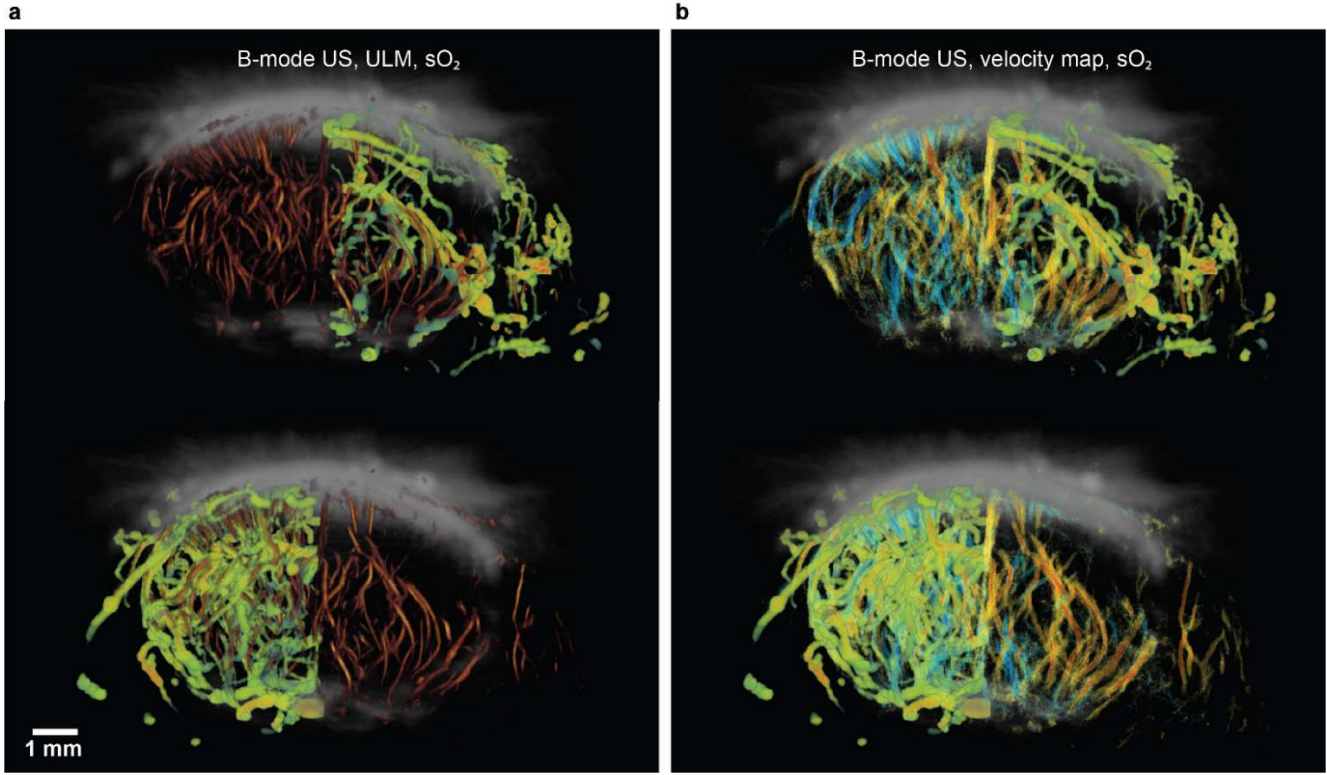

**Fig. 7. Co-registration of multi-modality images. (a)** Co-registered US B-mode, ULM vessel density, and PA $sO_2$ images. **(b)** Co-registered US B-mode, ULM velocity map, and PA $sO_2$ images. $sO_2$ of individual vessels were extracted using the segmented PD vessels as the mask.

Compared to the PD images, the 3D ULM images reveal more detailed vessel structures with an approximately 4–6-fold higher spatial resolution. For example, the spatial resolutions measured on two representative diving vessels were 269.1 μm and 444.9 μm in the PD image, and 64.3 μm and 68.1 μm in the ULM image, respectively (**Figs. 6e-g**). Nevertheless, the PD images have better vessel continuity than the ULM images, mainly because the ULM images need to accumulate a large number of microbubble trajectories.

*D. Muti-model Imaging Fusion and Quantitative Analysis*

Taking advantage of the complementary contrast and inherent co-registration of the PA, US B-mode, PD, and ULM images, we have explored the multi-modal imaging fusion, which can provide more comprehensive understanding of the brain functions than each single imaging alone. For example, by merging the PA $sO_2$ image with the ULM images (**Fig. 7**), we can observe that: 1) the lack of blood perfusion into the stroke region leads to the hypoxia in the stroke core; 2) the relatively high blood flow rate surrounding the stroke region is closely related to apparent high oxygenation level in the same region.

Acute stroke results in irreversible damage to the ischemic core, while the surrounding tissue known as the penumbra can potentially be salvaged. Without reperfusion, the tissue in the penumbra region undergoes necrosis, leading to an expansion of the infarct [31]. In TTC results, the white areas correspond to the ischemic core, whereas the pink rim surrounding it represents the penumbra region. Existing literature demonstrates that the ischemic region affected by pMCAO undergoes constant changes following the surgery. Over a period of 24 to 72 hours, the ischemic volume significantly increases by approximately 40%, indicating the progression of the stroke [32]. Therefore, regions displaying a lack of blood perfusion on PD and ULM images, while still exhibiting relatively high $sO_2$, are possibly penumbra regions. In these tissues, tissue viability is sustained, and there hasn't been sufficient time for the clots to become fully deoxygenated.

Moreover, using the vessel segmentation result from the PD image as the spatial mask, we can quantify various morphological and functional parameters on the same vessels from the PA and ULM images (**Table I**). For example, the total number of vessels in the stroke (right) and healthy (left) hemispheres are 162 and 231 (**Fig. 8a**), respectively, showing a 30% stroke-induced reduction in the vessel density. We also analyzed vessel radius, vessel length, $sO_2$, and flow speed of the two hemispheres, as shown in **Fig. 8b**. We fitted the histograms with a kernel density estimation to estimate the probability density function of vessel distribution in the healthy and stroke hemisphere. As expected, all of the quantified parameters showed significant differences in the stroke region. We performed two-sample Z-test between the two hemispheres and we observed that 1) the stroke hemisphere had a significant increase in mean vessel length and a reduction in vessel radius compared to the healthy hemisphere, suggesting that the small vessels were most affected by the stroke; 2) The vessels that were still perfused in the stroke hemisphere exhibited an ~7% decrease in $sO_2$ compared to those in the healthy hemisphere; and 3) The vessels in the stroke hemisphere had a median flow speed less than 4.6 mm/s, while the median speed for the healthy hemisphere was 5.2 mm/s.

**Table I.** Statistical analysis of the blood vessels extracted from stroke and healthy hemispheres.

| Parameter | Healthy (n = 219) | | |
|---|---|---|---|
| | Mean$_{healthy}$ | std$_{healthy}$ | Median |
| Radius (mm) | 0.099 | 0.029 | 0.092 |
| Length (mm) | 3.533 | 2.753 | 2.720 |
| $sO_2$ | 0.780 | 0.076 | 0.792 |
| Avg. speed (mm/s) | 5.198 | 1.266 | 5.149 |
| Parameter | Stroke (n = 134) | | |
| | Mean$_{stroke}$ | std$_{stroke}$ | Median |
| Radius (mm) | 0.090 | 0.026 | 0.081 |
| Length (mm) | 4.867 | 4.044 | 3.660 |
| $sO_2$ | 0.703 | 0.108 | 0.721 |
| Avg. blood flow speed (mm/s) | 4.577 | 1.296 | 4.571 |
| Parameter | Two-sample Z-test | | |
| | Mean$_{healthy}$ - Mean$_{stroke}$ | Z-test statistic | p-value |
| Radius (mm) | 0.009 | 3.12 | 0.00181 |
| Length (mm) | -1.335 | -3.69 | 0.000228 |
| $sO_2$ | 0.077 | 7.85 | 4.07E-15 |
| Avg. blood flow speed (mm/s) | 0.621 | 4.43 | 9.41E-06 |

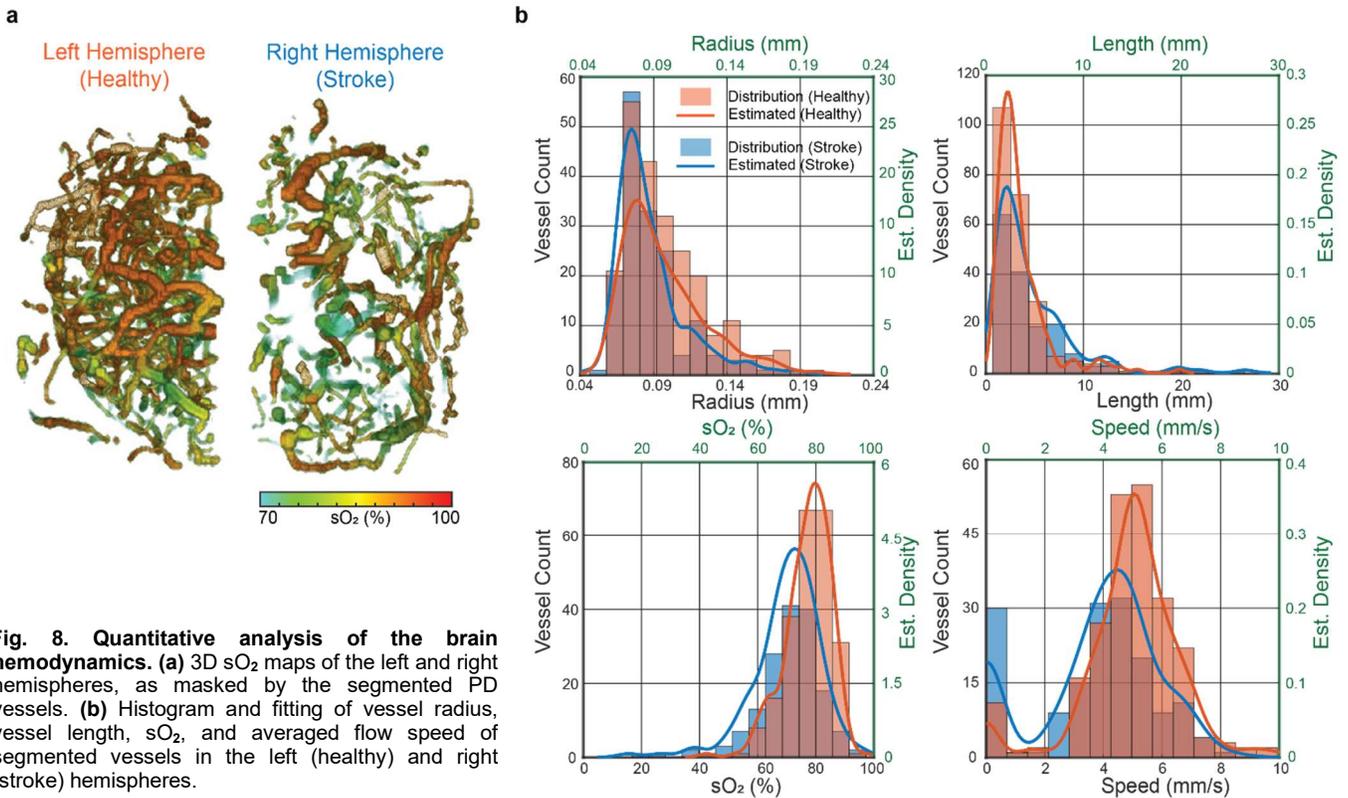

**Fig. 8. Quantitative analysis of the brain hemodynamics. (a)** 3D $sO_2$ maps of the left and right hemispheres, as masked by the segmented PD vessels. **(b)** Histogram and fitting of vessel radius, vessel length, $sO_2$, and averaged flow speed of segmented vessels in the left (healthy) and right (stroke) hemispheres.

## IV. Discussion

We have developed an integrated multi-modality imaging system, 3D-PAULM, that is inherently capable of both multi-spectral PA imaging and high-speed US imaging with complimentary contrasts. The system utilizes a 2D spherical transducer array at a relatively low frequency of 4 MHz and can provide whole-brain imaging with intact skin and skull. The system can achieve synthetic-aperture-based US imaging with a volumetric frame rate of 215 Hz and is capable of 3D PD imaging and ULM. To overcome the small focal zone of the transducer array, we stitched multiple locations for both PA and US imaging, and the entire imaging process with the extended field of view took ~10 minutes. We used the segmented vessels from the PD image as a mask to extract individual vessel information, including vasculature structure, $sO_2$, and blood flow speed.

One notable feature of our system is that it has achieved 3D ULM on the mouse brain with the skin and skull intact, which has always been a technical challenge for previous ULM methods. The key difference is that our imaging system uses a low-frequency transducer as opposed to the high-frequency ones typically employed. High-frequency ultrasound waves provide detailed images but are quickly attenuated by skull and tissue, thus limiting their depth of penetration. To compensate for potential resolution loss associated with lower frequencies, we utilized bandpass filtering and directional filtering for better microbubble localization [25]. This makes our system suitable for potential clinical applications where the imaging depth is critical. To further improve the accuracy of ULM, the volumetric frame rate of the US imaging should be improved, for example, by optimizing the transmission pattern with fewer transmitting events. Nevertheless, further studies are warranted to assess the impact of these changes on the ULM image quality. Moreover, the 3D PA imaging still has the limited view problem (e.g., the missing diving vessels). We can potentially address this issue by rotating the transducer array or by using microbubbles as virtual point sources [33].

The current image reconstruction method is 3D DAS with a homogenous speed of sound (SOS). The image quality and subsequent processing accuracy can be further improved by adopting multi-SOS reconstruction [34]. This can be achieved by mapping the skull location and thickness using the US B-mode image, as well as estimating the speed of sound heterogeneity. The use of more accurate reconstruction methods is expected to enhance PD imaging, and more significantly, the ULM imaging, as microbubbles can be more accurately reconstructed and better matched with the system's point spread

10 IEEE TRANSACTIONS ON MEDICAL IMAGING, VOL. XX, NO. XX, XXXX 2020function (PSF). Further improvement in localization accuracy can be achieved by incorporating the spatial variability of the system's PSF, for example, through deep-learning-based methods. In our study, vessel segmentation was only performed on the PD images, as vessel continuity is important for accurate vessel extraction. However, more volumetric frames can be acquired for complete microbubble localization, and vessel segmentation may be performed on the ULM images.

Since our system is totally non-invasive and capable of real-time imaging, it is possible to be applied for treatment guidance, such as monitoring thermal therapy of brain cancers [35], [36]. In addition, we can incorporate PA molecular imaging into our system, such as using photoswitchable phytochromes [37], [38]. With the combination of these inherently compatible image modalities, we believe that our technology can be a powerful tool for a broad range of preclinical and clinical studies.

## REFERENCES

[1] D. Wu, L. Huang, M. S. Jiang, and H. Jiang, "Contrast Agents for Photoacoustic and Thermoacoustic Imaging: A Review," *International Journal of Molecular Sciences*, vol. 15, no. 12, Art. no. 12, Dec. 2014,

[2] J. Xia, J. Yao, and L. V. Wang, "Photoacoustic tomography: principles and advances," *Electromagn Waves (Camb)*, vol. 147, pp. 1–22, 2014.

[3] V. Ntziachristos and D. Razansky, "Optical and Opto-Acoustic Imaging," in *Molecular Imaging in Oncology*, Springer, Berlin, Heidelberg, 2013, pp. 133–150.

[4] P. Beard, "Biomedical photoacoustic imaging," *Interface Focus*, Aug. 2011, Accessed: Sep. 20, 2019. [Online]. Available: https://royalsocietypublishing.org/doi/abs/10.1098/rsfs.2011.0028

[5] G. Paltauf, R. Nuster, M. Haltmeier, and P. Burgholzer, "Experimental evaluation of reconstruction algorithms for limited view photoacoustic tomography with line detectors," *Inverse Problems*, vol. 23, no. 6, pp. S81–S94, Nov. 2007,

[6] S. C. Hester, M. Kuriakose, C. D. Nguyen, and S. Mallidi, "Role of Ultrasound and Photoacoustic Imaging in Photodynamic Therapy for Cancer," *Photochemistry and Photobiology*, vol. 96, no. 2, pp. 260–279, Mar. 2020,

[7] M. B. Karmacharya, L. R. Sultan, and C. M. Sehgal, "Photoacoustic monitoring of oxygenation changes induced by therapeutic ultrasound in murine hepatocellular carcinoma," *Sci Rep*, vol. 11, no. 1, Art. no. 1, Feb. 2021,

[8] G. S. Sangha and C. J. Goergen, "Label-free photoacoustic and ultrasound imaging for murine atherosclerosis characterization," *APL Bioeng*, vol. 4, no. 2, Jun. 2020,

[9] Y. Zhang and L. Wang, "Video-Rate Ring-Array Ultrasound and Photoacoustic Tomography," *IEEE Transactions on Medical Imaging*, vol. 39, no. 12, pp. 4369–4375, Dec. 2020,

[10] C. Errico et al., "Ultrafast ultrasound localization microscopy for deep super-resolution vascular imaging," *Nature*, vol. 527, no. 7579, pp. 499–502, Nov. 2015,

[11] O. Couture, V. Hingot, B. Heiles, P. Muleki-Seya, and M. Tanter, "Ultrasound Localization Microscopy and Super-Resolution: A State of the Art," *IEEE Transactions on Ultrasonics, Ferroelectrics, and Frequency Control*, vol. 65, no. 8, pp. 1304–1320, Aug. 2018,

[12] N. Renaudin, C. Demené, A. Dizeux, N. Ialy-Radio, S. Pezet, and M. Tanter, "Functional ultrasound localization microscopy reveals brain-wide neurovascular activity on a microscopic scale," *Nat Methods*, vol. 19, no. 8, Art. no. 8, Aug. 2022,

[13] E. Stride and N. Saffari, "Microbubble ultrasound contrast agents: A review," *Proc Inst Mech Eng H*, vol. 217, no. 6, pp. 429–447, Jun. 2003,

[14] E. Quaia, "Microbubble ultrasound contrast agents: an update," *Eur Radiol*, vol. 17, no. 8, pp. 1995–2008, Aug. 2007,

[15] J. Robin, A. Özbek, M. Reiss, X. L. Dean-Ben, and D. Razansky, "Dual-Mode Volumetric Optoacoustic and Contrast Enhanced Ultrasound Imaging With Spherical Matrix Arrays," *IEEE Transactions on Medical Imaging*, vol. 41, no. 4, pp. 846–856, Apr. 2022,

[16] S. Zhao, J. Hartanto, R. Joseph, C.-H. Wu, Y. Zhao, and Y.-S. Chen, "Hybrid photoacoustic and fast super-resolution ultrasound imaging," *Nat Commun*, vol. 14, no. 1, Art. no. 1, Apr. 2023,

[17] J. A. Jensen, S. I. Nikolov, K. L. Gammelmark, and M. H. Pedersen, "Synthetic aperture ultrasound imaging," *Ultrasonics*, vol. 44, pp. e5–e15, Dec. 2006,

[18] B. E. Treeby and B. T. Cox, "k-Wave: MATLAB toolbox for the simulation and reconstruction of photoacoustic wave fields," *JBO*, vol. 15, no. 2, p. 021314, Mar. 2010,

[19] L. Menozzi, Á. del Águila, T. Vu, C. Ma, W. Yang, and J. Yao, "Three-dimensional non-invasive brain imaging of ischemic stroke by integrated photoacoustic, ultrasound and angiographic tomography (PAUSAT)," *Photoacoustics*, vol. 29, p. 100444, Feb. 2023,

[20] B. Kelechava, "ANSI Z136.1-2022: Safe Use of Lasers - ANSI Blog," *The ANSI Blog*, Mar. 13, 2023. https://blog.ansi.org/ansi-z136-1-2022-safe-use-of-lasers/ (accessed Jul. 14, 2023).

[21] M. Li, Y. Tang, and J. Yao, "Photoacoustic tomography of blood oxygenation: A mini review," *Photoacoustics*, vol. 10, pp. 65–73, Jun. 2018,

[22] B. Cox, J. G. Laufer, S. R. Arridge, and P. C. Beard, "Quantitative spectroscopic photoacoustic imaging: a review," *J Biomed Opt*, vol. 17, no. 6, p. 061202, Jun. 2012,

[23] P. Song et al., "Improved Super-Resolution Ultrasound Microvessel Imaging With Spatiotemporal Nonlocal Means Filtering and Bipartite Graph-Based Microbubble Tracking," *IEEE Transactions on Ultrasonics, Ferroelectrics, and Frequency Control*, vol. 65, no. 2, pp. 149–167, Feb. 2018,

[24] C. Demené et al., "Spatiotemporal Clutter Filtering of Ultrafast Ultrasound Data Highly Increases Doppler and fUltrasound Sensitivity," *IEEE Transactions on Medical Imaging*, vol. 34, no. 11, pp. 2271–2285, Nov. 2015,

[25] C. Huang et al., "Short Acquisition Time Super-Resolution Ultrasound Microvessel Imaging via Microbubble Separation," *Sci Rep*, vol. 10, no. 1, Art. no. 1, Apr. 2020,

[26] B. Heiles et al., "Volumetric Ultrasound Localization Microscopy of the Whole Rat Brain Microvasculature," *IEEE Open Journal of Ultrasonics, Ferroelectrics, and Frequency Control*, vol. 2, pp. 261–282, 2022,

[27] K. Jaqaman et al., "Robust single-particle tracking in live-cell time-lapse sequences," *Nat Methods*, vol. 5, no. 8, Art. no. 8, Aug. 2008,

[28] P. Klippel, "Enabling fast image reconstruction for photoacoustic tomography using GPUs," 2022.

[29] S. R. Aylward and E. Bullitt, "Initialization, noise, singularities, and scale in height ridge traversal for tubular object centerline extraction," *IEEE Transactions on Medical Imaging*, vol. 21, no. 2, pp. 61–75, Feb. 2002,

[30] R. W. Regenhardt et al., "Blood Pressure and Penumbral Sustenance in Stroke from Large Vessel Occlusion," *Frontiers in Neurology*, vol. 8, 2017.

[31] S. Liu, S. R. Levine, and H. R. Winn, "Targeting ischemic penumbra: part I - from pathophysiology to therapeutic strategy," *J Exp Stroke Transl Med*, vol. 3, no. 1, pp. 47–55, Mar. 2010.

[32] N. C. Shanbhag, R. H. Henning, and L. Schilling, "Long-term survival in permanent middle cerebral artery occlusion: a model of malignant stroke in rats," *Sci Rep*, vol. 6, no. 1, Art. no. 1, Jun. 2016,

[33] Y. Tang et al., "High-fidelity deep functional photoacoustic tomography enhanced by virtual point sources," *Photoacoustics*, vol. 29, p. 100450, Feb. 2023,

[34] M. Cui, H. Zuo, X. Wang, K. Deng, J. Luo, and C. Ma, "Adaptive photoacoustic computed tomography," *Photoacoustics*, vol. 21, p. 100223, Mar. 2021,

[35] E. A. Gonzalez, A. Jain, and M. A. L. Bell, "Combined Ultrasound and Photoacoustic Image Guidance of Spinal Pedicle Cannulation Demonstrated With Intact ex vivo Specimens," *IEEE Transactions on Biomedical Engineering*, vol. 68, no. 8, pp. 2479–2489, Aug. 2021,

[36] H. Song et al., "Real-Time Intraoperative Surgical Guidance System in the da Vinci Surgical Robot Based on Transrectal Ultrasound/Photoacoustic Imaging With Photoacoustic Markers: An Ex Vivo Demonstration," *IEEE Robotics and Automation Letters*, vol. 8, no. 3, pp. 1287–1294, Mar. 2023,

[37] J. Yao et al., "Multiscale photoacoustic tomography using reversibly switchable bacterial phytochrome as a near-infrared photochromic probe," *Nat Methods*, vol. 13, no. 1, Art. no. 1, Jan. 2016,

[38] L. Li et al., "Small near-infrared photochromic protein for photoacoustic multi-contrast imaging and detection of protein interactions in vivo," *Nat Commun*, vol. 9, no. 1, Art. no. 1, Jul. 2018,